\documentclass[11pt]{article}
\usepackage{latexsym, amssymb}
\usepackage{graphics}
\title{Non-Existence of Black Hole Solutions to Static, Spherically
Symmetric Einstein-Dirac Systems -- a Critical Discussion}
\author{Felix Finster, Joel Smoller, and Shing-Tung Yau}
\date{November 2002}

\newfont{\Bb}{msbm10 scaled 1095}

\newcommand\R{{\mbox{\Bb R}}}

\begin{document}
\maketitle
\begin{abstract}
This short note compares different methods to prove that Einstein-Dirac
systems have no static, spherically symmetric solutions.
\end{abstract}
There are several approaches to prove that Einstein-Dirac systems
do not admit static, spherically symmetric solutions. The first
paper in this direction is~\cite{FSY1}, where it is shown that the
Dirac equation has no normalizable, time-periodic solutions in the
Reissner-Nordstr{\"o}m geometry. In~\cite{FSY2}--\cite{FSY4} non-existence
results were
obtained for coupled static systems by choosing polar coordinates
and analyzing the nonlinear radial ODEs. Recently, M.\ Dafermos~\cite{D}
proposed a different method where he analyzes the Einstein
equations in null coordinates. His method has the advantage that
it also applies to other Einstein-matter systems and
generalizes to time-periodic solutions.

The mathematical and physical assumptions under which the above methods
apply are quite different. Therefore, these methods are not
equivalent, and it is rather subtle to decide which approach
is preferable for a given physical system.
The purpose of this short note is to compare the different approaches by
collecting and discussing the necessary assumptions and the obtained results.

The physical situation of interest is the spherically symmetric
collapse to a black hole. Thus thinking of the Cauchy problem for
a coupled Einstein-matter system, we consider an initial Cauchy
surface which is topologically $\R^3$, such that an event horizon forms
in its future development. It is a reasonable physical assumption
that asymptotically for large time, the system should settle down
to a static (or more generally time-periodic) system. Under this
assumption, ruling out non-trivial static solutions means that the
matter (as described by the Dirac field) is no longer present
asymptotically as $t \to \infty$, and thus the
matter must either have fallen into the black hole or must have
escaped to infinity.
For this physical interpretation to hold, it is
essential that the assumptions, under which the non-existence
result for the static Einstein-Dirac system applies, are satisfied
in the gravitational collapse.

\begin{figure}[t]
\begin{center}
{\includegraphics{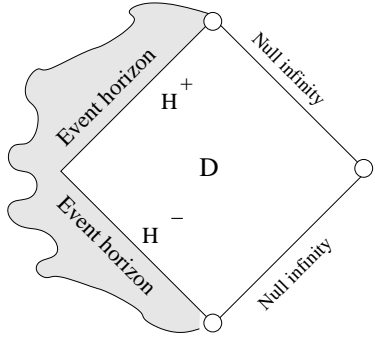}}
\caption{Conformal diagram of the extended space-time}
\label{aplo}
\end{center}
\end{figure}

We now discuss the individual papers in chronological order.
In the first paper~\cite{FSY1} the situation is particularly simple
in that the gravitational field is a given Reissner-Nordstr{\"o}m
background field. In the non-extreme case,
the problem is analyzed in the maximally extended Kruskal space-time,
where the domain of outer communications $D$ is connected to both
a black hole and a white hole through the event horizons $H_+$ and
$H_-$, respectively (see the conformal diagram in Figure~\ref{aplo};
the figures are taken from~\cite{D}).
In the context of the gravitational collapse described above,
the Reissner-Nordstr{\"o}m metric should be considered as
physical space-time only asymptotically as $t \to \infty$.
In particular, the physical metric should involve a black hole,
but no white hole. In order to take this physical input into account in the
time-periodic situation, it was imposed that the integral of
the Dirac current over the event horizon of the white hole $H_-$
be zero. This assumption means that no matter should come out of
the white hole, corresponding to the fact that in the physical
space-time no white hole is present.

The physical picture of the gravitational collapse also tells us that
one must be very careful in imposing regularity conditions for
the wave function $\Psi$ near the event horizons.
First of all, we recall that due to current conservation,
the probability integral
\[ \int_{\cal{H}} \overline{\Psi} G^k \Psi\: \nu_k\: d\mu_{\cal{H}} \]
is independent of the choice of the hypersurface~${\cal{H}}$.
Here the integrand (also called the ``probability density'')
is positive when the hypersurface is spacelike, but it may
be negative when the hypersurface is non-spacelike. This causes
problems when we consider the probability integral over the
hypersurface $t={\mbox{const}}$ in Schwarzschild coordinates
up to the event horizon $r=r_1$. Namely, the hypersurface  $t={\mbox{const}}$
is spacelike outside and non-spacelike inside the event horizon.
The gravitational collapse tells us that the probability integral
over the hypersurface $t={\mbox{const}}$ should be finite when we integrate
across the horizon. But since the integrand may be negative inside
the horizon, there seems no reason why the probability integrals
inside and outside the event horizons should separately be finite.
Because of this difficulty, in~\cite{FSY1} it was assumed merely
that the probability integral is finite {\em{outside and away from the
event horizon}}, i.e.\ when we integrate from $r=r_1+\varepsilon$ to $r=\infty$.
Transforming a plane wave $\sim e^{i \omega t}$ to Kruskal coordinates,
one sees furthermore that functions which are smooth
in Schwarzschild coordinates will in Kruskal coordinates in general be highly
singular near $H^+$ and $H^-$. This regularity problem is bypassed
in~\cite{FSY1} by evaluating the Dirac equation in a neighborhood of
the event horizons {\em{only weakly}}. This gives rise to
so-called matching conditions. Using that the Dirac current is
divergence-free, we finally show that the matching conditions are
incompatible with the finite probability integral outside and away of
the event horizon unless $\Psi$ vanishes identically (= radial flux argument).
In the extreme case, a completely different method involving a barrier
argument for the spinors is used, but we shall not discuss this here.

In the subsequent papers~\cite{FSY2}--\cite{FSY4} various
static Einstein-Dirac systems are analyzed.
Because of the coupling to matter, nothing is known about the asymptotic
form of the metric near the event horizon. Therefore, we cannot extend
the metric across the event horizon, and it is impossible to derive
matching conditions.
For this reason, the solutions are considered only outside the event
horizon. It is convenient to choose polar coordinates $(t,r,\vartheta,\varphi)$
where the problem reduces to analyzing the radial ODEs.
As in~\cite{FSY1}, it is merely assumed that the
probability integral should be zero outside and away of the event horizon.
Furthermore, it is assumed that the volume element $\sqrt{|\det g_{ij}|}$
is smooth and non-zero on the event horizon; this means physically
that an observer who is freely falling into the black hole should not
feel strong tidal forces when crossing the horizon. Finally,
a regularity assumption is needed for the metric
function $A(r)$ near the event horizon.
More precisely, in~\cite{FSY2} and~\cite{FSY4} it is assumed that
$A$ has near the event horizon a power expansion of the form
$A(r) \sim (r-r_1)^\kappa + o((r-r_1)^\kappa)$ with $\kappa>0$.
In~\cite{FSY3} only monotonicity of $A$ near $r=r_1$ is needed.
The regularity assumption on $A$ has no physical motivation, it is
a purely technical assumption. On the other
hand, this assumption is very weak and seems to cover
all cases of physical interest.

\begin{figure}[t]
\begin{center}
{\includegraphics{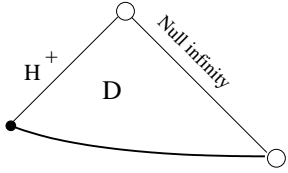}}
\caption{Conformal diagram of physical space-time}
\label{prin}
\end{center}
\end{figure}

We finally discuss the approach by Dafermos~\cite{D}. His interesting methods
make it possible to treat general classes of spherical systems
that are either static or time-periodic.
In contrast to~\cite{FSY2}--\cite{FSY4}, where the equations
are considered in polar coordinates, Dafermos uses null coordinates.
He also has gravitational collapse in mind and thus assumes
that physical space-time has a conformal diagram as shown in
Figure~\ref{prin}.
As explained above, the physical picture is
that space-time should go over to a static space-time
asymptotically near time-like infinity.
Dafermos needs that this static space-time has a conformal diagram
with two intersecting event horizons $H^\pm$ as
shown in Figure~\ref{aplo}.
This assumption is clearly satisfied in the non-extreme
Reissner-Nordstr{{\"o}}m metric (as considered in~\cite{FSY1}).
But for a coupled Einstein-Dirac
system, there seems no reason why this assumption should be satisfied.
In particular, physical arguments do not apply because physical
space-time is static only asymptotically near time-like infinity,
and thus it is unphysical to consider the static space-time for all
(also negative) times.
The assumption that static space-time has two intersecting
event horizon implies that
the event horizon is of the same type as in the non-extreme
Reissner-Nordstr{{\"o}}m geometry (i.e.~that a neighborhood of the origin in the
conformal diagram in Figure~\ref{aplo} is diffeomorphic to
the neighborhood of the origin in Kruskal coordinates).
This is a very strong assumption, which excludes in particular
the power behavior $A(r) \sim (r-r_1)^\kappa$, $\kappa \neq 1$, as considered in~\cite{FSY2}.
Moreover, Dafermos imposes very strong regularity assumptions
by assuming that the Dirac current $\overline{\Psi} G^\alpha \Psi$ is locally
$C^1$. This implies in particular that the probability integral
is finite if one integrates up to the event horizon.
As discussed above, this too is a strong assumption.
Finally, the assumption~$\overline{\Psi} G^\alpha \Psi \in C^1_{\mbox{\scriptsize{loc}}}$
is also problematic because, as explained above,
even smooth functions in polar coordinates will in general be very
singular on $H^\pm$.

To summarize, the approach~\cite{D} has the advantage that it
applies to a general class of systems and to time-periodic solutions.
However, when specialized to Einstein-Dirac systems,
it requires strong assumptions
on the causal structure of static space-time and on the regularity
of the wave function $\Psi$. In physical situations where these strong assumptions
do not obviously hold, the methods in~\cite{FSY1}--\cite{FSY4} are preferable.
All methods have in common that particular
properties of the Dirac current $\overline{\Psi} G^\alpha \Psi$ are used: it
is a divergence-free, future-directed time-like vector field.
Thus the current conservation and the positivity of the probability density
are a key ingredient to all non-existence proofs.

\begin{tabular}{lll}
\\
Felix Finster & Joel Smoller &  Shing-Tung Yau \\
NWF I -- Mathematik & Mathematics Department
& Mathematics Department \\
Universit{{\"a}}t Regensburg & The University of Michigan &
Harvard University \\
93040 Regensburg, Germany & Ann Arbor, MI 48109, USA & Cambridge, MA 02138,
USA \\
{\tt{Felix.Finster@mathematik}} &
{\tt{smoller@umich.edu}} & {\tt{yau@math.harvard.edu}} \\
$\;\;\;\;\;\;\;\;\;\;\;\:${\tt{.uni-regensburg.de}} & &
\end{tabular}


\begin{thebibliography}{99}

\bibitem{D} M.\ Dafermos, ``On `time-periodic' black-hole solutions
to certain spherically symmetric Einstein-matter systems,''
gr-qc/0210025

\bibitem{FSY1} F.\ Finster, J.\ Smoller, S.-T.\ Yau,
``Non-existence of time-periodic solutions of the Dirac
equation in a Reissner-Nordstr{\"o}m black hole background,''
gr-qc/9805050, J.\ Math.\ Phys.\ {\bf 41} (2000) 2173--2194

\bibitem{FSY2} F.\ Finster, J.\ Smoller, S.-T.\ Yau,
``Non-existence of black hole solutions for a spherically symmetric,
static Einstein-Dirac-Maxwell system,'' gr-qc/9810048,
Comm.\ Math.\ Phys.\ {\bf 205} (1999), no. 2, 249--262

\bibitem{FSY3} F.\ Finster, J.\ Smoller, S.-T.\ Yau,
``The interaction of Dirac particles with non-abelian
gauge fields and gravity--black holes,'' gr-qc/9910047,
Mich.\ Math.\ J.\ {\bf 47} (2000) 199-208

\bibitem{FSY4} F.\ Finster, J.\ Smoller, S.-T.\ Yau,
``Absence of static, spherically symmetric black hole solutions for
Einstein-Dirac-Yang/Mills Equations with complete fermion shells,''
gr-qc/0005028, Adv.\ Theor.\ Math.\ Phys.\ {\bf{4}} (2000) 1231-1257

\end{thebibliography}
\end{document}